\documentclass[10pt]{article}
\usepackage{graphicx}
\usepackage[T1]{fontenc}

\usepackage[utf8]{inputenc}
\usepackage{amsmath}
\usepackage{amsfonts}
\usepackage{amssymb}
\usepackage{calrsfs}
\usepackage[mathscr]{euscript}

\usepackage{float}

\def\Title#1{\begin{center} {\Large #1 } \end{center}}
\def\Author#1{\begin{center}{ \sc #1} \end{center}}
\def\Address#1{\begin{center}{ \it #1} \end{center}}

\newcommand\pubblock{\rightline{\begin{tabular}{l} Proceedings of the Fifth Annual LHCP\\ \pubnumber\\
         \pubdate  \end{tabular}}}

\newenvironment{Abstract}{\begin{quotation} \begin{center} 
             \large ABSTRACT \end{center}\bigskip 
      \begin{center}\begin{large}}{\end{large}\end{center} \end{quotation}}

\newenvironment{Presented}{\begin{quotation} \begin{center} 
             PRESENTED AT\end{center}\bigskip 
      \begin{center}\begin{large}}{\end{large}\end{center} \end{quotation}}

\def\beq{\begin{equation}}
\def\eeq#1{\label{#1}\end{equation}}
\def\eeqn{\end{equation}}

\def\beqa{\begin{eqnarray}}
\def\eeqa#1{\label{#1}\end{eqnarray}}
\def\eeqan{\end{eqnarray}}

\let\bar=\overbar

\def\Dslash{\not{\hbox{\kern-4pt $D$}}}
\def\dslash{\not{\hbox{\kern-2pt $\del$}}}

\def\msb{{\bar{\ssstyle M \kern -1pt S}}}

\usepackage{color}

\newcommand\pubnumber{ ATL-PHYS-PROC-2017-XXX }

\newcommand\pubdate{\today}

\def\affiliation{
On behalf of the CMS Collaboration \\
Department of Physics \\
University and INFN, Torino, Italy}

\begin{document}

\large
\begin{titlepage}
\pubblock

\vfill
\Title{  Higgs properties measurements using the four lepton decay channel  }
\vfill

\Author{ Muhammad Bilal Kiani  }
\Address{\affiliation}
\vfill
\begin{Abstract}

The measurements of the properties of the Higgs boson are presented in the H$\rightarrow$ZZ$\rightarrow$4$\ell$ ($\ell$=e,$\mu$) decay channel using a data sample corresponding to an integrated luminosity of 35.9 fb$^{-1}$ of proton-proton collisions at a center-of-mass energy of 13 TeV recorded by the CMS detector at the LHC. The signal-strength modifier $\mu$, defined as the production cross section of the Higgs boson times its branching fraction to four leptons relative to the standard model expectation, is measured to be $\mu=1.05^{+0.19}_{-0.17}$ at $m_{\mathrm{H}}=125.09~\mathrm{GeV}$. Constraints are set on the strength modifiers for the main Higgs boson production modes. The mass is measured to be $m_{\mathrm{H}}=125.26 \pm 0.21~\mathrm{GeV}$ and the width is constrained using on-shell production to be $\Gamma_{\mathrm{H}}<1.10~\mathrm{GeV}$, at $95\%$ CL. The fiducial cross section is measured to be $2.90^{+0.48}_{-0.44}({\rm stat.})^{+0.27}_{-0.22}({\rm sys.})~{\mathrm{fb}}$, which is compatible with the standard model prediction of $2.72\pm0.14~{\mathrm{fb}}$. 

\end{Abstract}
\vfill

\begin{Presented}
The Fifth Annual Conference\\
 on Large Hadron Collider Physics \\
Shanghai Jiao Tong University, Shanghai, China\\ 
May 15-20, 2017
\end{Presented}
\vfill
\end{titlepage}
\def\thefootnote{\fnsymbol{footnote}}
\setcounter{footnote}{0}
%

\normalsize 


\section{Introduction}

The H$\rightarrow$ZZ$\rightarrow 4\ell $  decay channel ($\ell = e,\mu$) has a large signal-to-background ratio due to the complete reconstruction of the final state decay products and excellent lepton momentum resolution. This makes it one of the most important channels for studies of the Higgs boson's properties. Measurements performed using this decay channel and the Run 1 data set include the determination of the mass and spin-parity of the new boson, its width and fiducial cross sections, as well as tests for anomalous HVV couplings. \cite{Chatrchyan:2013mxa}

The results are presented on the measurements of properties of the Higgs boson in the H$\rightarrow$ZZ$\rightarrow 4\ell$ decay channel at $\sqrt{s}=13$TeV. Categories have been introduced targeting subleading production modes of the Higgs boson such as vector boson fusion (VBF) and associated production with a vector boson (WH, ZH) or top quark pair ttH. In addition, dedicated measurements of the boson's mass, width, total and differential cross sections have been presented.

\section{Event Selection}

The  Z candidates are formed with pairs of leptons of the same flavor and opposite-charge ($e^{+} e^{-}$, $\mu^{+}\mu^{-}$) and required to pass 12 < $m_{\ell^{+}\ell^{-}}$  < 120GeV.
They are then combined into ZZ candidates, wherein we denote as Z$_{1}$ the Z candidate with an invariant mass closest to the nominal Z boson mass, and as Z$_{2}$ the other one. The flavors of involved leptons define three mutually exclusive subchannels: $4e$, $4\mu$ and $2e 2\mu$.

To be considered for the analysis, ZZ candidates have to pass a set of kinematic requirements that improve the sensitivity to Higgs boson decays.
The Z$_{1}$ invariant mass must be larger than $40$GeV.
All leptons must be separated in angular space by at least $\Delta R(\ell_i, \ell_j) > 0.02$.
At least two leptons are required to have p$_{T} > 10$ GeV at least one is required to have p$_{T}$ > 20GeV.

To further suppress events with leptons originating from hadron decays in jet fragmentation or from the decay of low-mass hadronic resonances, all four opposite-charge lepton pairs that can be built with the four leptons (irrespective of flavor) are required to satisfy $m_{\ell^{+}\ell^{-}}$ > 4GeV, where selected FSR photons are disregarded in the invariant mass computation.
Finally, the four-lepton invariant mass m$_{4\ell}$ must be larger than $70$GeV, which defines the mass range of interest for the subsequent steps of the analysis. The results of the events selection is shown in the Figure \ref{fig:ESC}(Left)

\begin{figure}[ht]
\centering
\begin{tabular}{ccc}
& \includegraphics[height=2in]{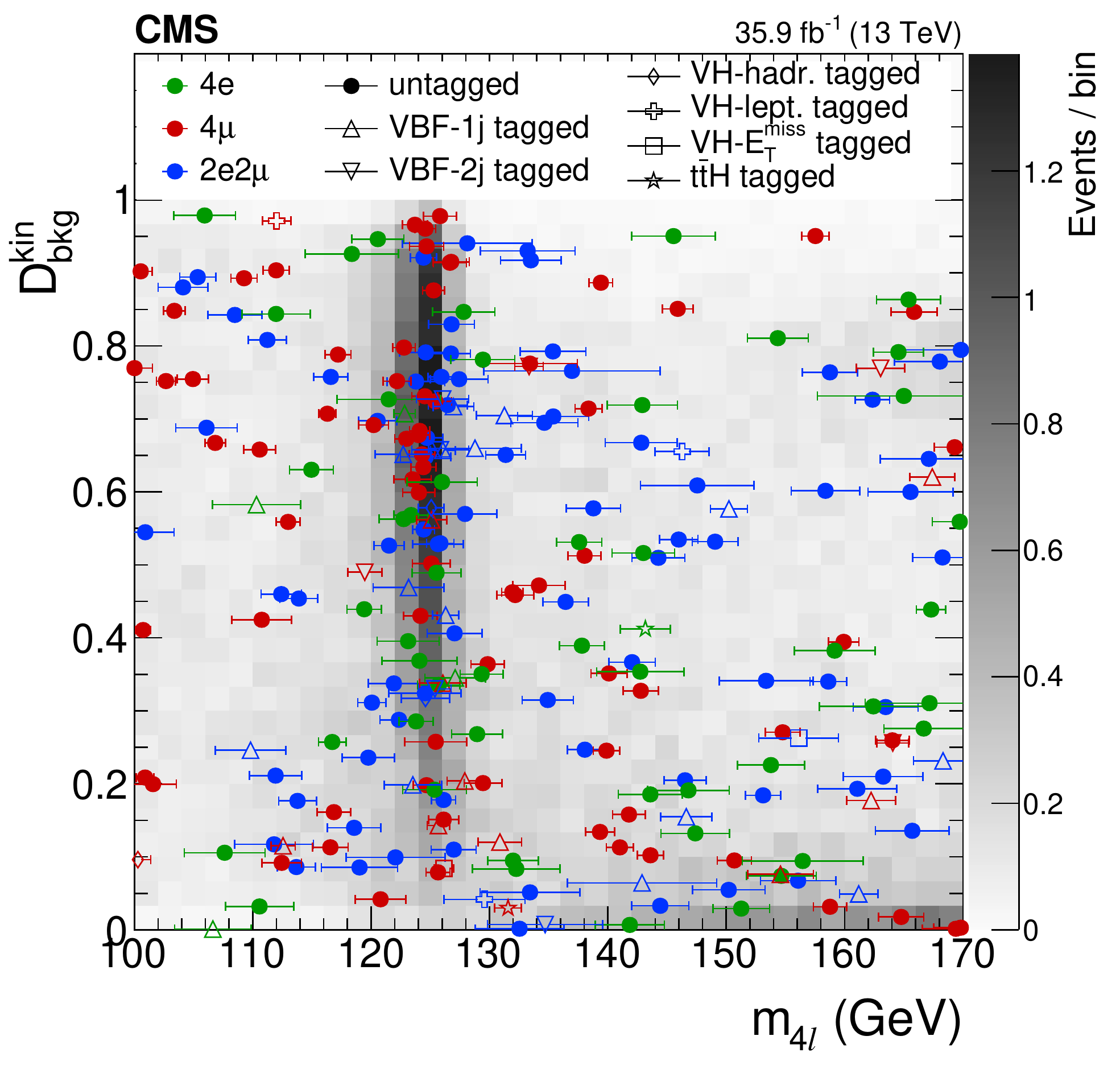} & \\
\includegraphics[height=1.5in]{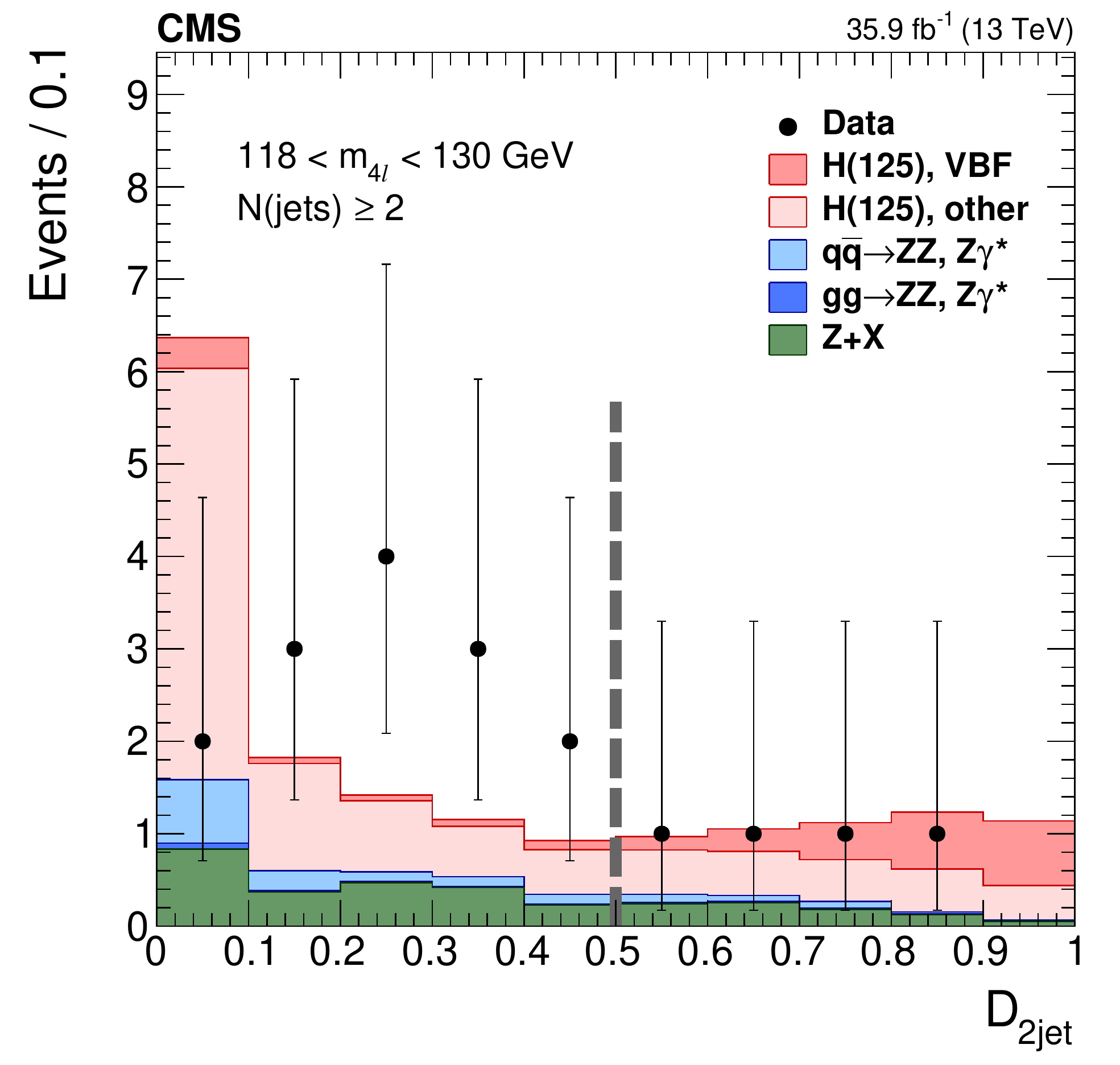} &  \includegraphics[height=1.5in]{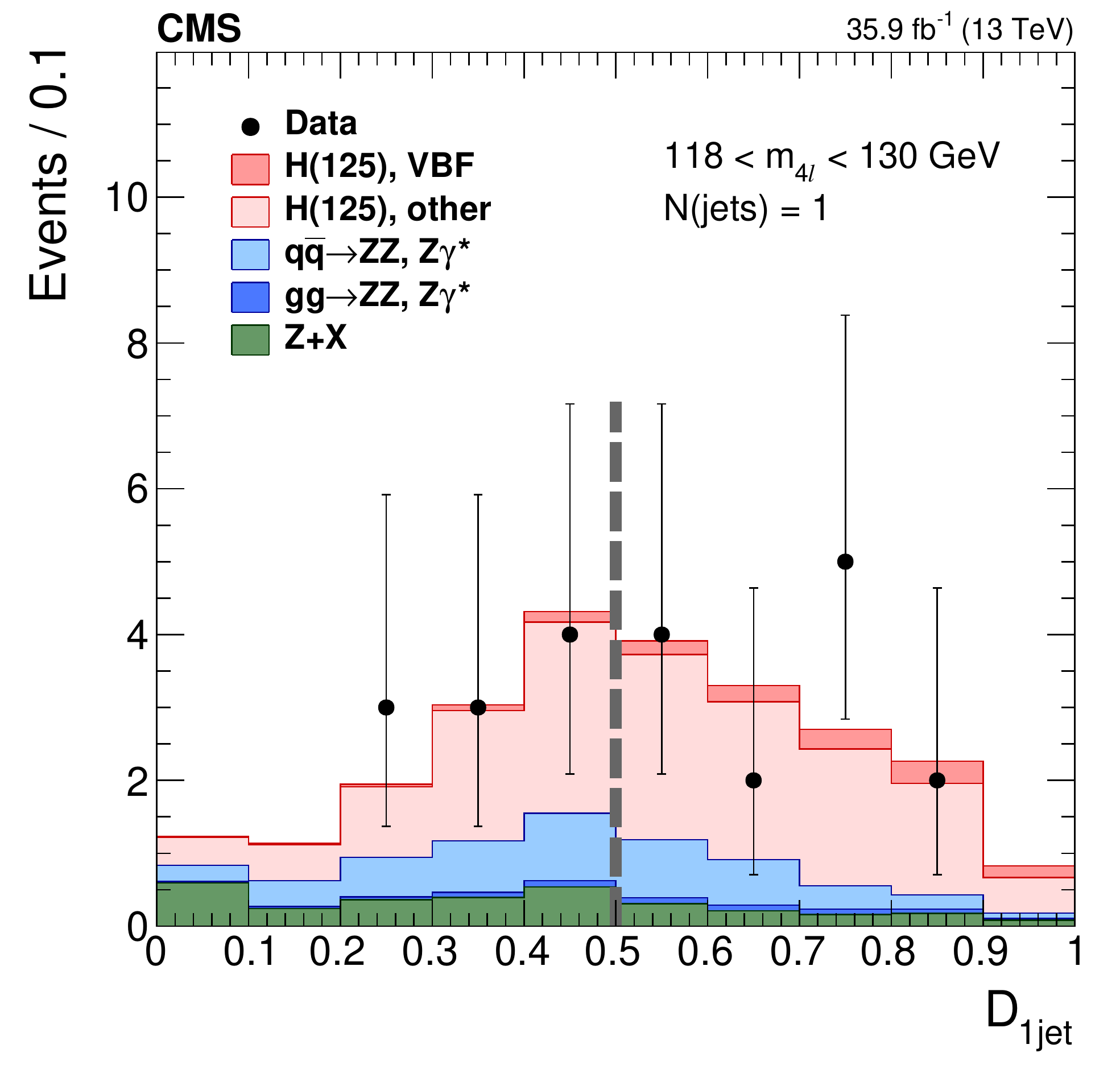} &
\includegraphics[height=1.5in]{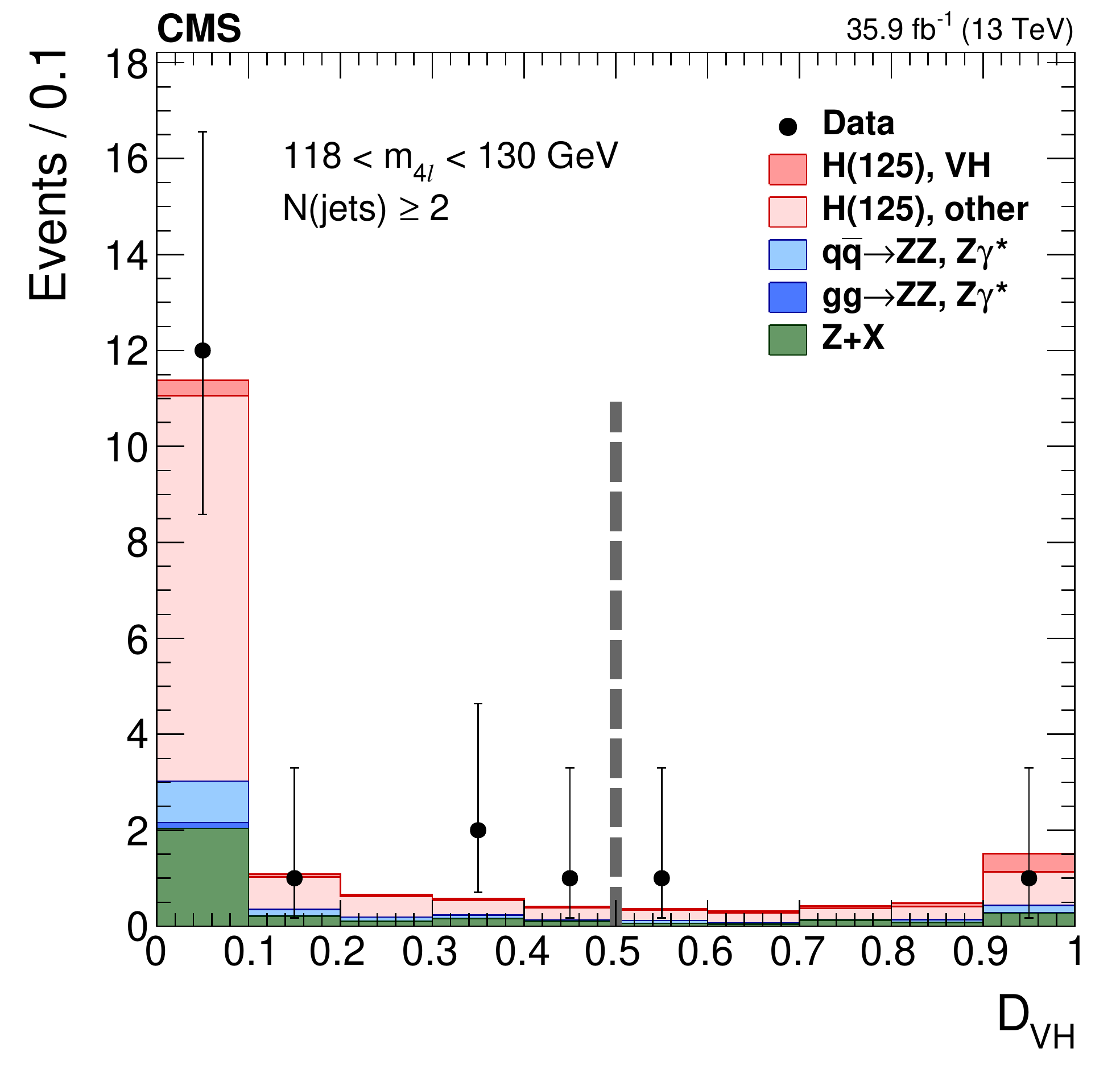}\\
\end{tabular}
\caption{ \label{fig:disc} Top: Distribution of $\ensuremath{\mathcal{D}^\text{kin}_\text{bkg}} $ versus $m_{4\ell}$   The gray scale represents the expected total number of ZZ background and SM Higgs boson signal events for $m_{H}$ = 125 GeV, Bottom: Distribution of categorization discriminants (Left) $\mathcal{D}_{2jet}$. (Middle) $\mathcal{D}_{1jet}$. (Right) $\mathcal{D}_{VH}$ = max($\mathcal{D}_{WH}$,$\mathcal{D}_{ZH}$) in Ref \cite{Sirunyan:2017exp}}
\end{figure}

\section{Event Categorization}
\begin{figure}[ht]
\centering
\begin{tabular}{cc}
\includegraphics[height=2in]{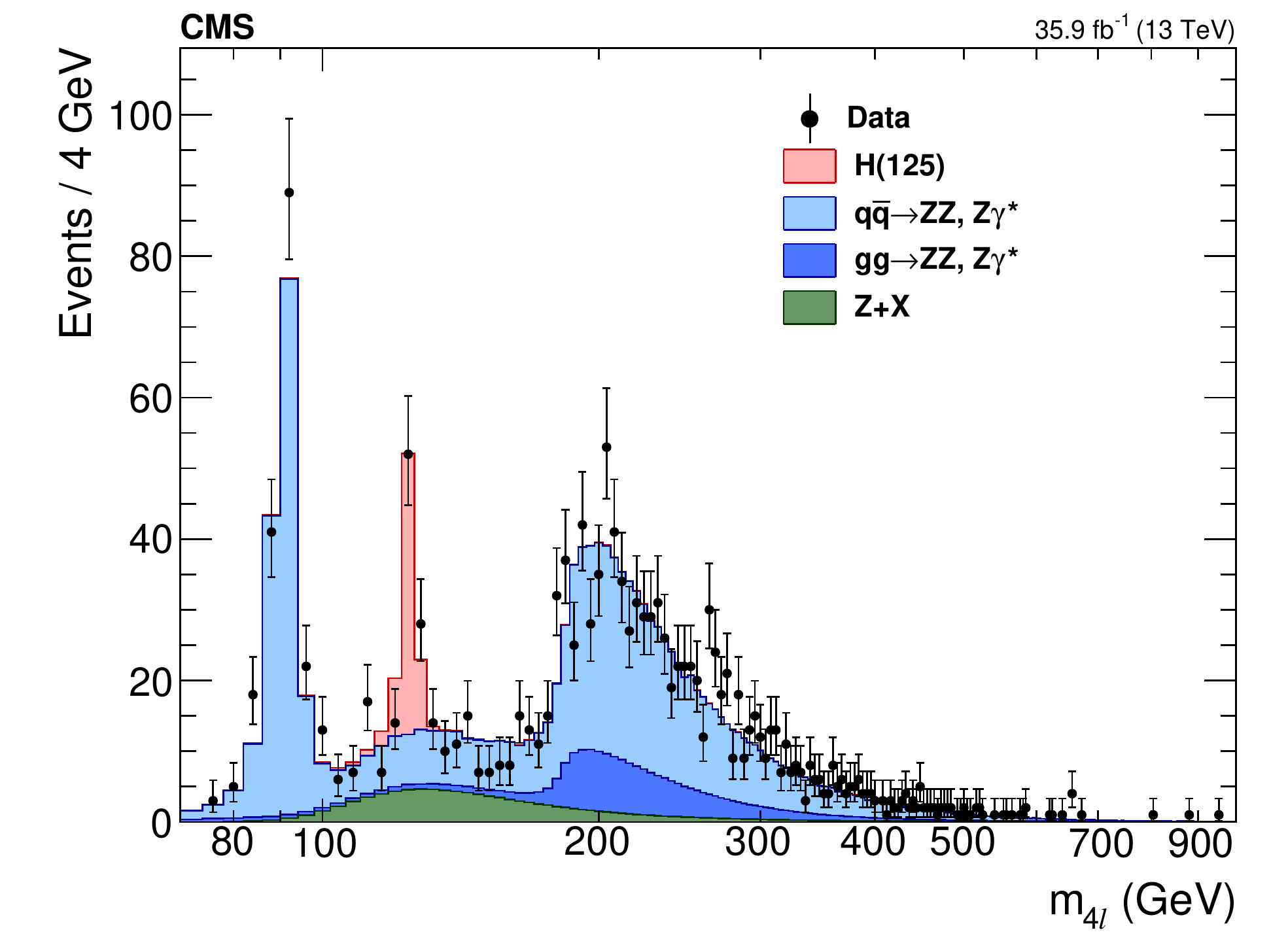} &
\includegraphics[height=2in]{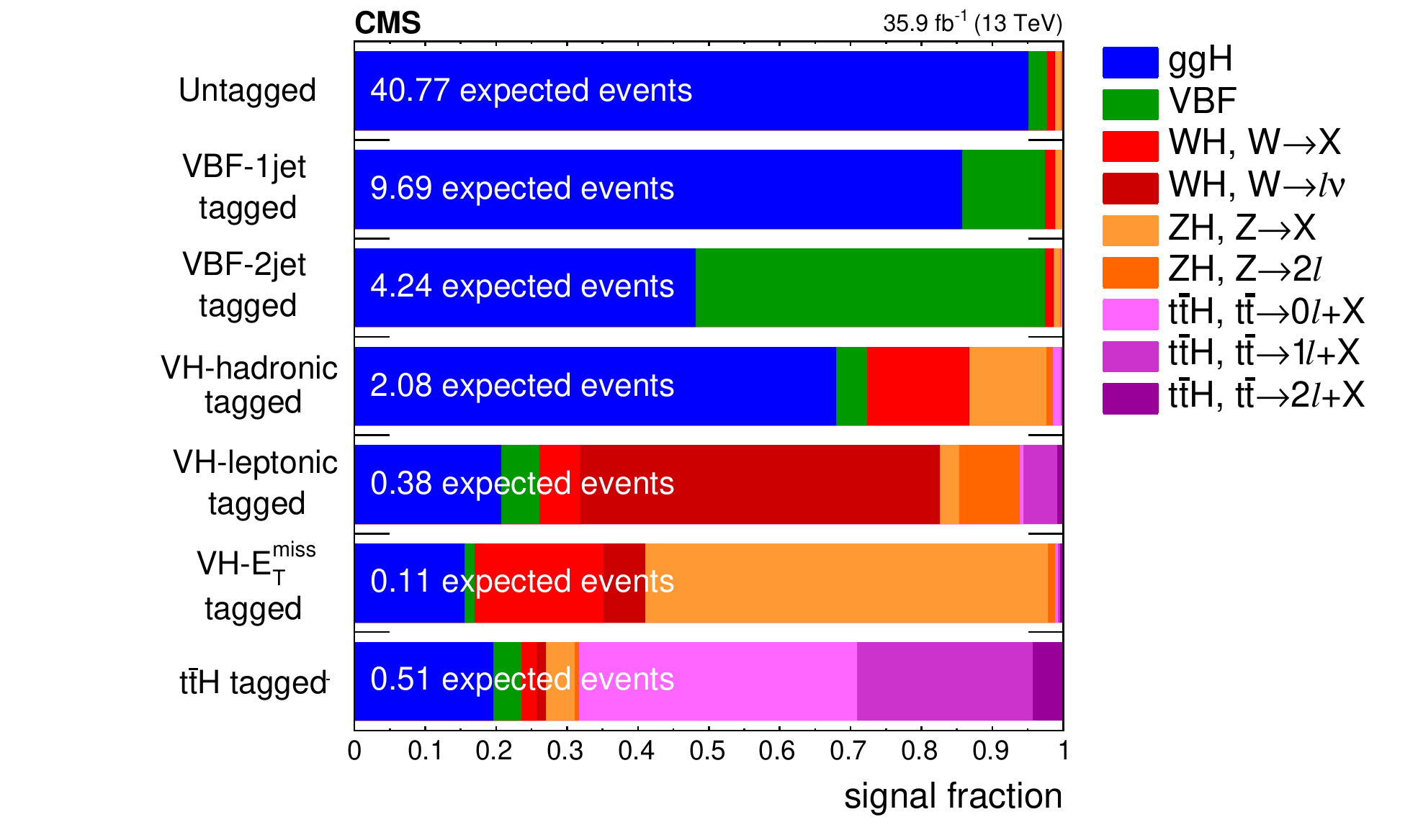} \\
\end{tabular}
\caption{ \label{fig:ESC}  Distribution of the four-lepton reconstructed invariant mass $m_{4\ell}$ in the full mass range (left)the signal relative purity of the seven event categories in terms of Higgs boson production processes(Right) in Ref \cite{Sirunyan:2017exp}}
\end{figure}

To improve the sensitivity to the Higgs boson production mechanisms, the selected events are classified into mutually exclusive categories. Seven categories are defined, using the following criteria applied in this exact order. Figure \ref{fig:ESC}(right) shows the signal relative purity of different production processes. 

The full kinematic information from each event using either the Higgs boson decay products or associated particles in its production is extracted using matrix element calculations and used to form several kinematic discriminants.  The discriminant sensitive to $gg/q\bar{q} \rightarrow 4\ell $  kinematics is $\ensuremath{\mathcal{D}^\text{kin}_\text{bkg}}$ and the discriminants $\mathcal{D}_{1jet}$,$\mathcal{D}_{2jet}$ and  $\mathcal{D}_{VH}$ = max($\mathcal{D}_{WH}$,$\mathcal{D}_{ZH}$) are used to target a specific Higgs production mode. The full definition of the observables can be found in Refs \cite{Khachatryan:2014kca} \cite{Spin} \cite{Spin1}.

\begin{itemize}
\item {\bf VBF-2jet-tagged}: exactly 4 leptons. In addition there must be either 2 or 3 jets of which at most 1 is b-tagged, or at least 4 jets and no b-tagged jets. Finally, ${\cal D}_{\rm 2jet}>0.5$ is required.
\item {\bf VH-hadronic-tagged}: exactly 4 leptons. In addition there must be 2 or 3 jets, or at least 4 jets and no b-tagged jets. Finally, ${\cal D}_{\rm VH} \equiv {\rm max}({\cal D}_{\rm ZH},{\cal D}_{\rm WH})>0.5$ is required.
\item {\bf VH-leptonic-tagged}: no more than 3 jets and no b-tagged jets in the event,
and exactly 1 additional lepton or 1 additional pair of opposite sign same flavor leptons. This category also includes events with no jets and at least 1 additional lepton.
\item {\bf ttH-tagged}: at least 4 jets of which at least 1 is b-tagged, or at least 1 additional lepton.
\item {\bf VH-MET-tagged}: exactly 4 leptons, no more than 1 jet and E$_T^{\rm miss}$ > 100GeV.
\item {\bf VBF-1jet-tagged}: exactly 4 leptons, exactly 1 jet and ${\cal D}_{\rm 1jet}>0.5$.
\item {\bf Untagged}: consists of the remaining events.
\end{itemize}

\section{Results}

\subsection*{Signal Strength}

\begin{figure}[ht]
\centering
\begin{tabular}{cc}
\includegraphics[height=2in]{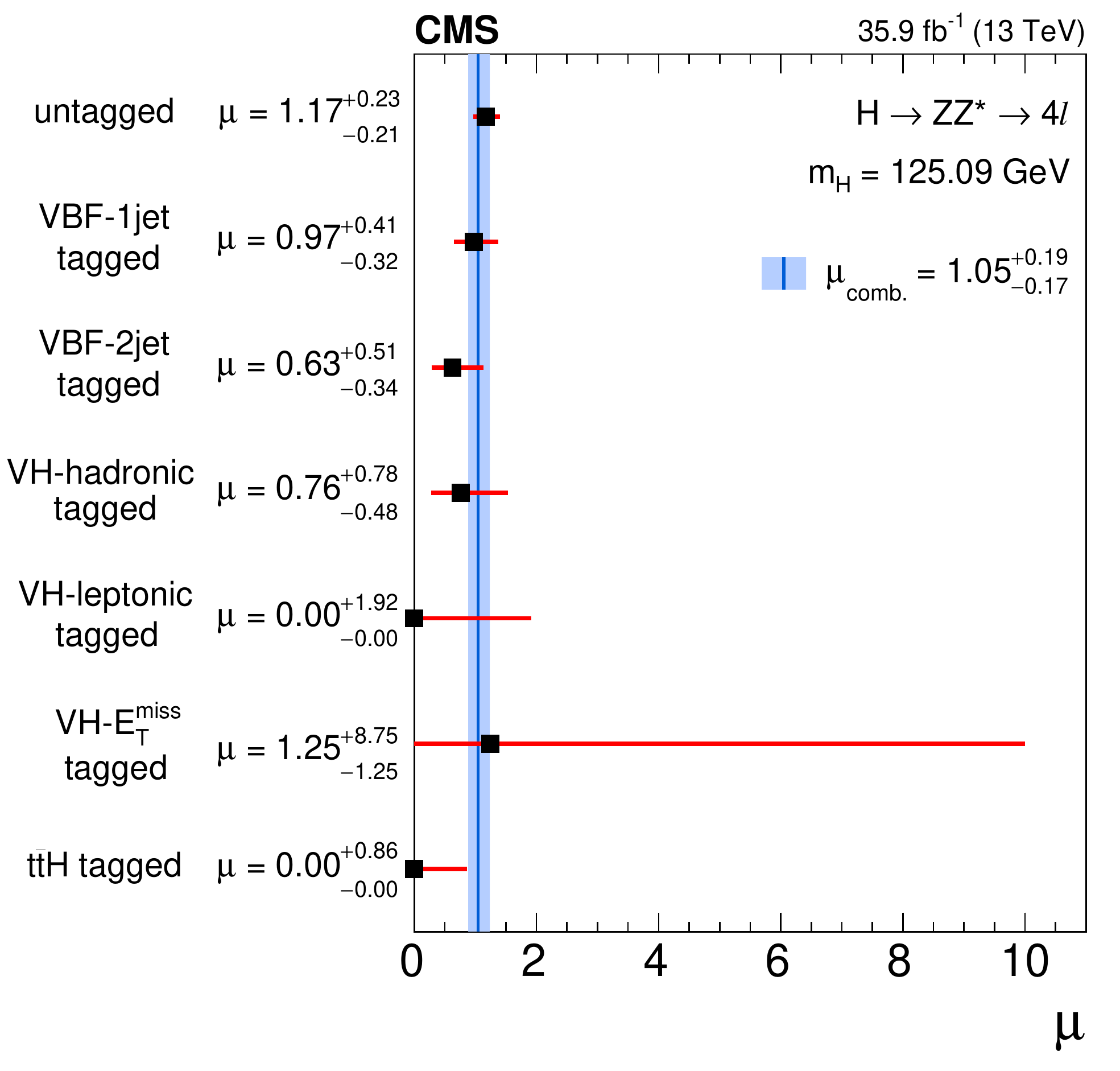} &
\includegraphics[height=2in]{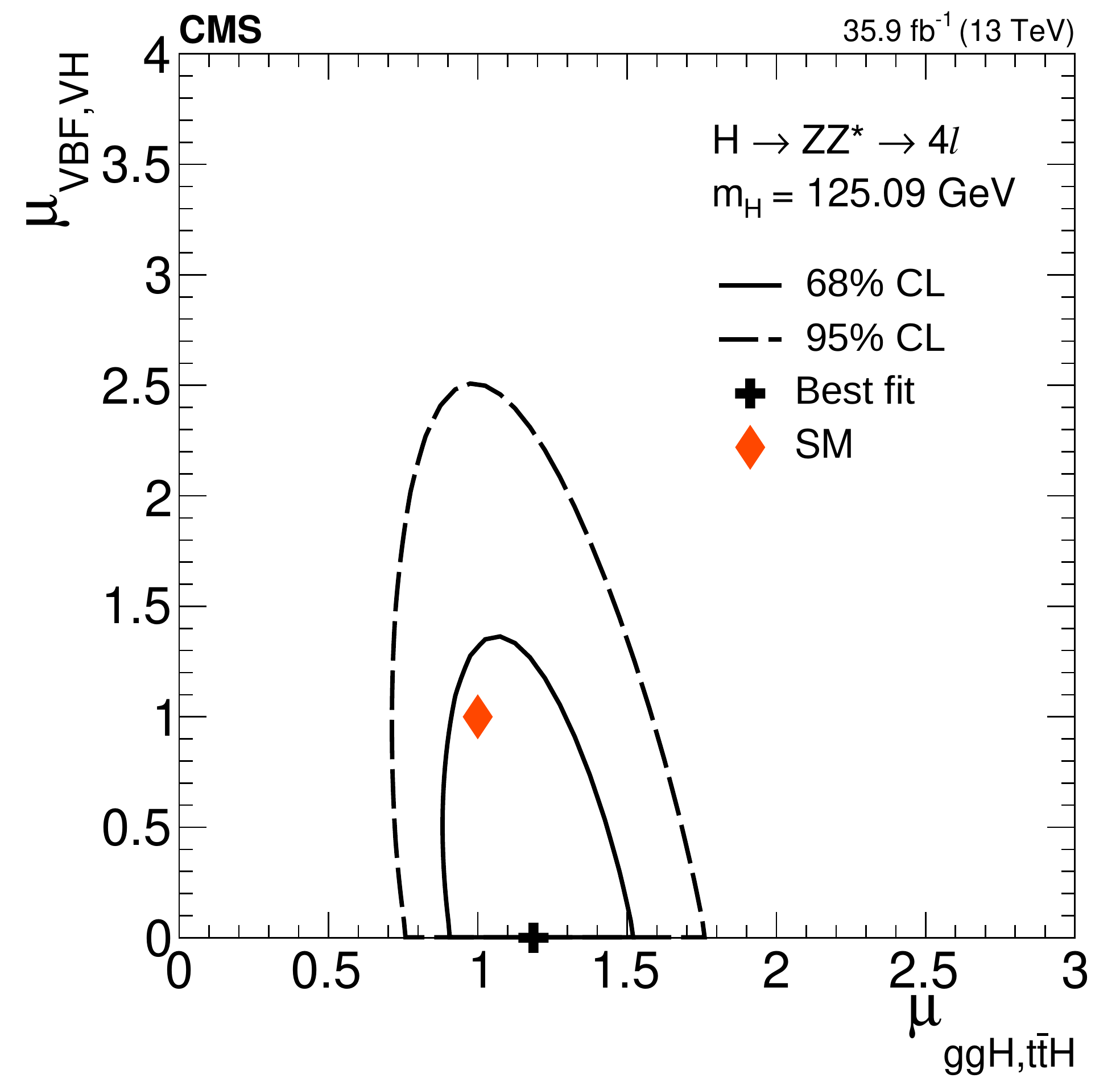} \\
\end{tabular}
\caption{ \label{fig:figure3} Observed values of the signal strength for the seven event categories, compared to the combined $\mu$ shown as a vertical line (left) Result of the 2D likelihood scan for the $\mu_{F}$ and $\mu_{V}$ signal strength modifiers. The solid and dashed contours show the 68\% and 95\% CL regions, respectively(Right) in Ref \cite{Sirunyan:2017exp}}
\end{figure}

The signal-strength is defined as the production cross section of the Higgs boson times its branching fraction to four leptons relative to  the standard model expectation. To extract the signal strength modifier we perform a multi-dimensional fit that relies on two variables in all the analysis categories: the four-lepton invariant mass $m_{4l}$ and the $\mathcal{D}^{\text{kin}}_{\text{bkg}}$ discriminant. We define the two-dimensional likelihood function as:

\begin{equation}
\mathcal{L}_{2D}(m_{4l},\mathcal{D}^\text{kin}_\text{bkg}) = \mathcal{L}(m_{4l}) \mathcal{L}(\mathcal{D}^\text{kin}_\text{bkg}|m_{4l}) .
\end{equation}
Figure \ref{fig:figure3} shows the results. 

\subsection*{Fiducial Cross Section}

\begin{figure}[ht]
\centering
\includegraphics[height=2.4in]{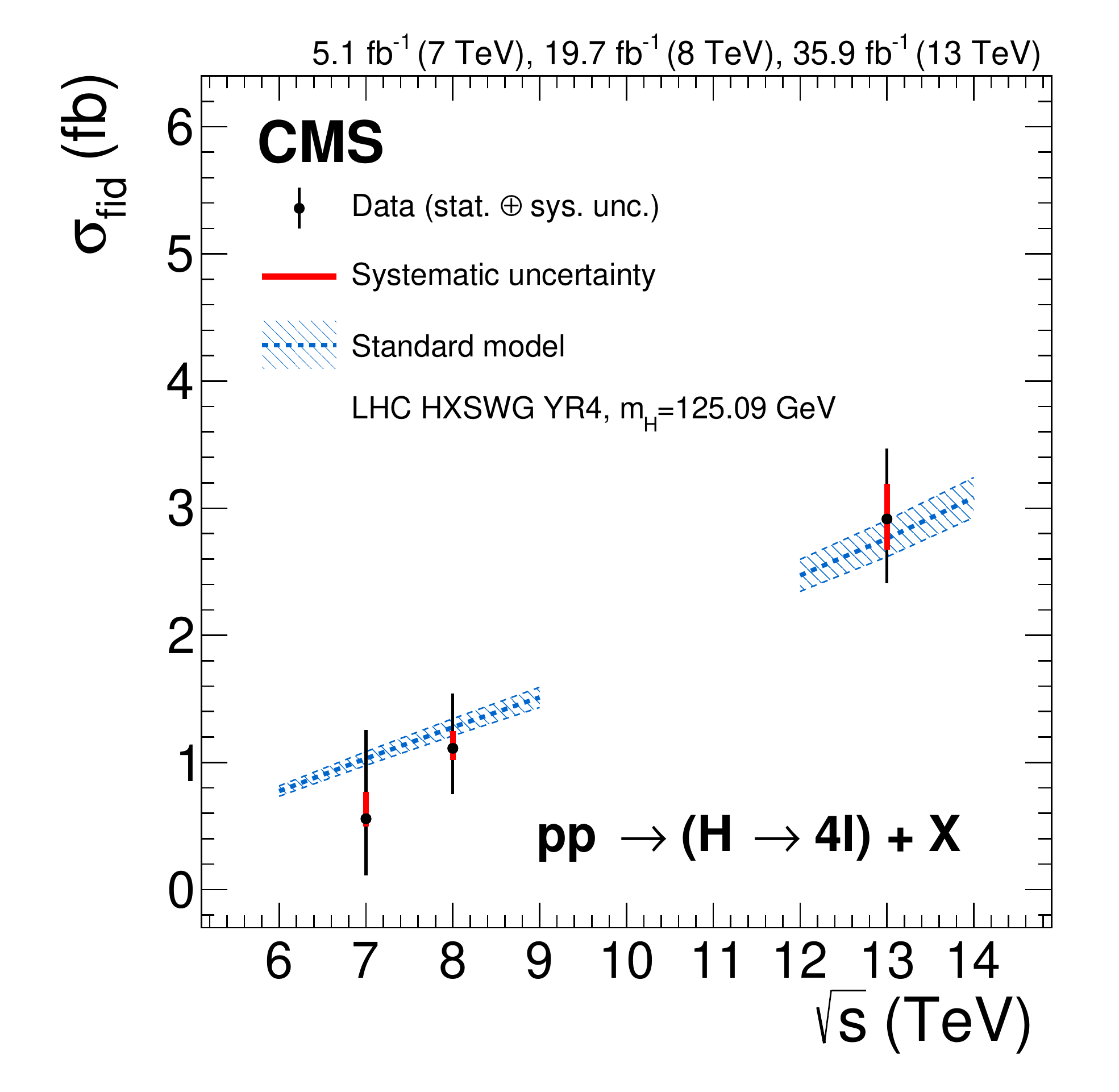} 
\caption{ \label{fig:figurexs} The measured fiducial cross section as a function of center of mass energy in Ref \cite{Sirunyan:2017exp} }
\end{figure}

The measurement of the cross section for the production and decay pp $\rightarrow$ H $\rightarrow$ 4$\ell$ within a fiducial volume defined to match closely the reconstruction level selection is presented. This measurement has minimal dependence on the assumptions of the relative fraction or kinematic distributions of the separate production modes. A maximum likelihood fit of the signal and background parameterizations to the observed 4$\ell$
mass distribution, is performed to extract the integrated fiducial cross section without categorization and use of discriminants. Figure \ref{fig:figurexs} shows the results for the fiducial cross section as a function of center of mass energy.

\subsection*{Mass and Width}
\begin{figure}[ht]
\centering
\begin{tabular}{cc}
\includegraphics[height=2in]{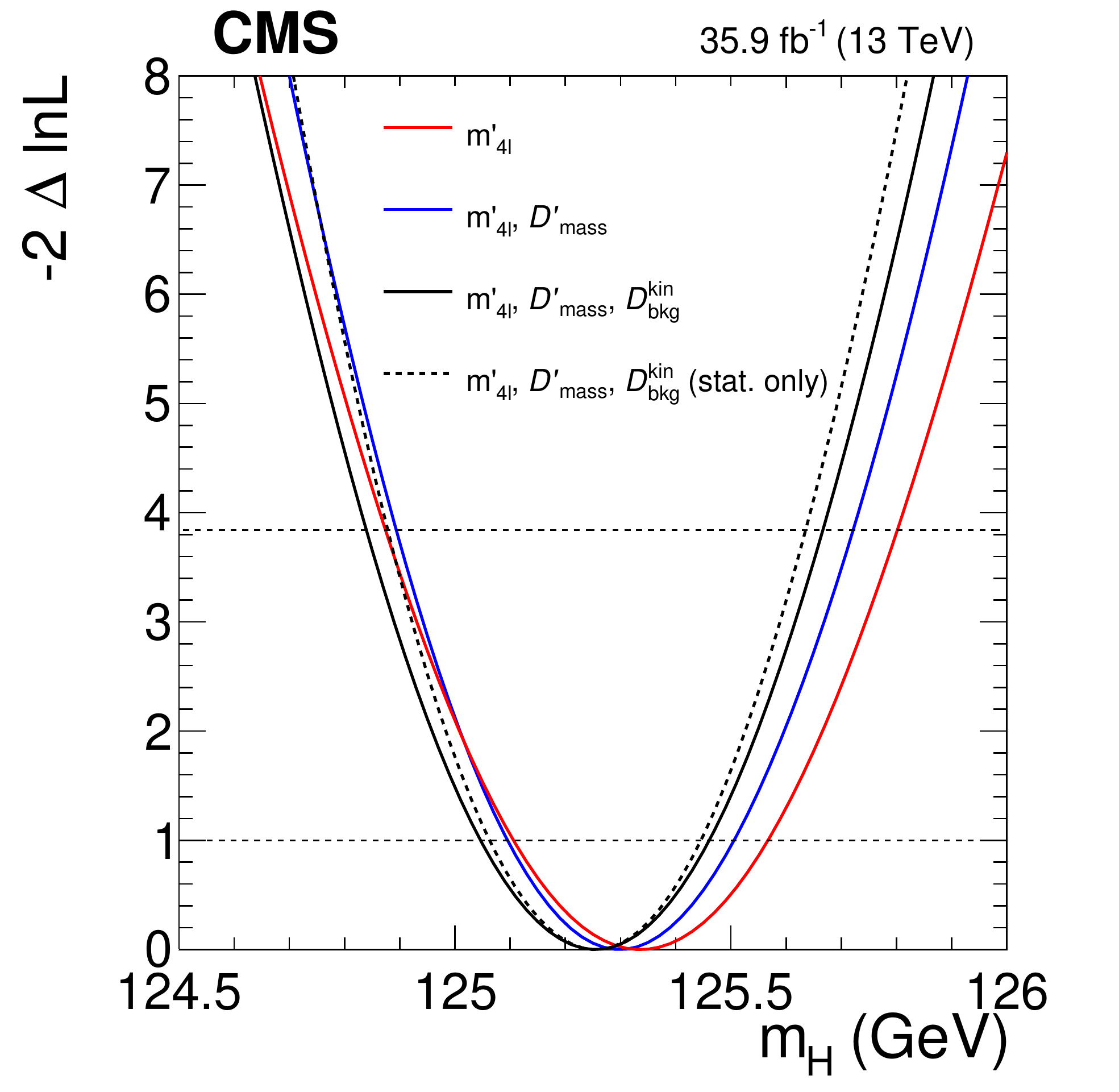} &
\includegraphics[height=2in]{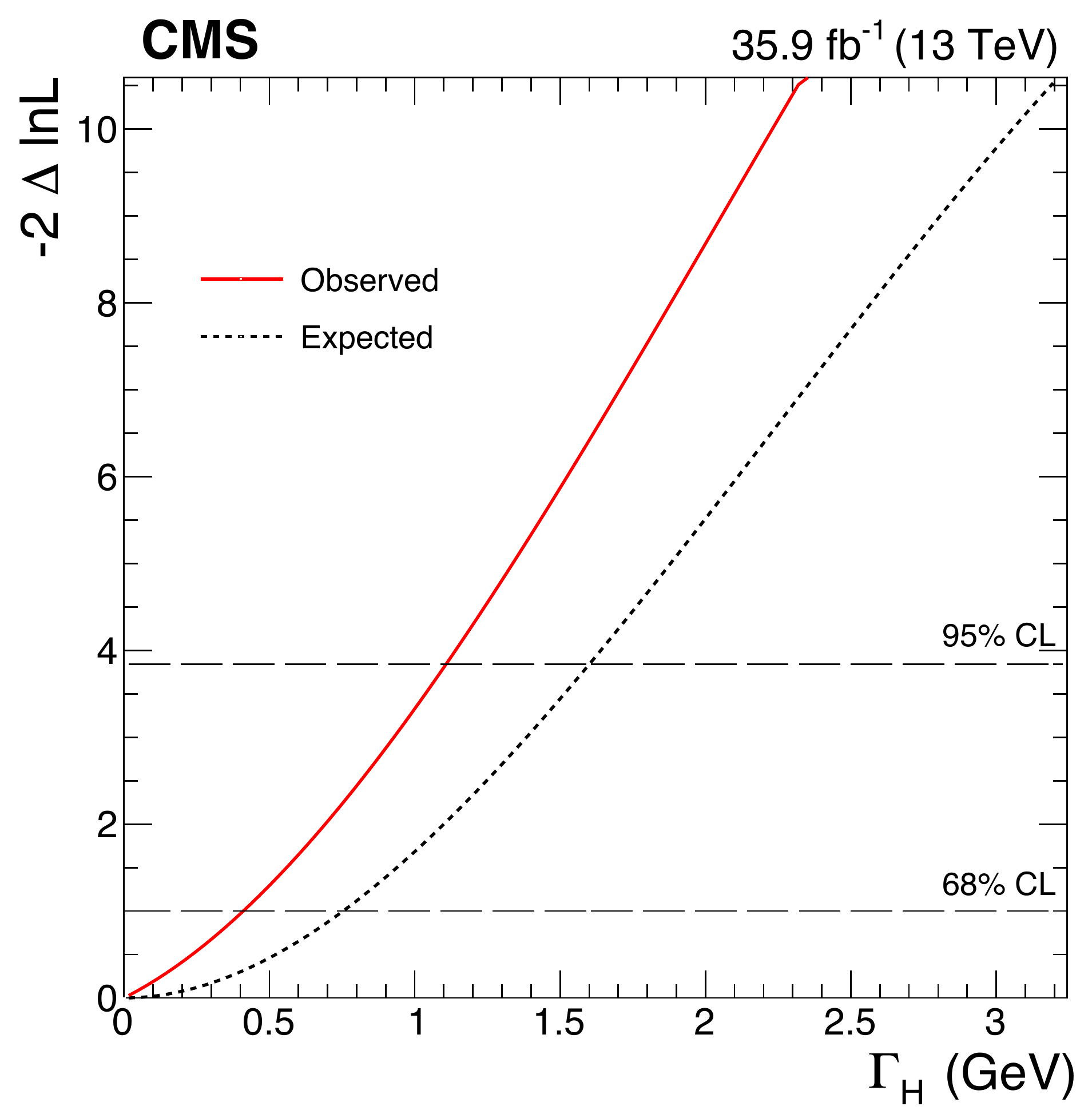} \\
\end{tabular}
\caption{ \label{fig:figureM} Left: 1D likelihood scan as a function of mass for the 1D, 2D, and 3D measurement,The likelihood scans are shown for the mass measurement using the refitted mass distribution with m(Z1) constraint, Right: Observed and expected likelihood scan of $\Gamma_{H}$ using the signal range 105 < $m_{4\ell}$ < 140 GeV, with $m_{H}$ floated in Ref \cite{Sirunyan:2017exp}}
\end{figure}

The measurement of the mass of the Higgs boson exploits additional information
from per-event relative mass uncertainties $\mathcal{D}_\text{mass}$, which are defined by propagating per-lepton momentum errors to the 4$\ell$ candidate. Using this variable brings an expected improvement of about 8$\%$ to the uncertainty of the mass measurement. Figure \ref{fig:figureM}(Left) shows the results for 1D,2D and 3D likelihood scan. 

A measurement of the width performed using on-shell Higgs boson production. An unbinned maximum likelihood fit to the $m_{4\ell}$ distribution is performed over the range of selected events. The strength of fermion-induced couplings and vector-boson-induced couplings are independent and are left unconstrained in the fit. For such a large width, interference between the signal and background production of the $m_{4\ell}$ final state becomes important and is taken into account in analysis. Results shown in Figure \ref{fig:figureM} (Right).

\section{Conclusions}
Several measurements of Higgs boson production in the four-lepton final state at $\sqrt{s}$ = 13TeV have been presented, using data samples corresponding to an integrated luminosity of 35.9fb$^{-1}$. All results are consistent, within their uncertainties, with the expectations for the SM Higgs boson. The detailed analysis is explained in the paper \cite{Sirunyan:2017exp}.

\end{document}